# Spin-Orbit Coupling Parameters of $sp^2$ Nanocarbons

Elena F. Sheka[1] and Elena V. Orlenko[2]

[1]Peoples' Friendship University of the Russia, Miklukho-Maklaya str. 6, 117198, Moscow, Russian Federation

[2]Peter the Great St.Petersburg Polytechnic University, Polytechnicheskaya, 29, 194021, St.Petersburg, Russian Federation

**Abstract**. The paper presents a new approach to the determination of standard spin-orbit coupling parameters, such as the SOC constant $a_{SO}$ and the Lande parameter g, by using peculiarities of UHF results for open-shell systems. The approach is convincingly approved for stretched ethylene and fullerene $C_{60}$ as examples, thus justifying at once a high efficacy of the UHF formalism to disclose SOC features.

## Introduction

Shown recently [1], spin-orbit coupling (SOC) is the main mechanism responsible for peculiar properties of open-shell molecules, such as spin-contamination of the ground state; lowering of the total energy; appearance of effectively unpaired electrons of $N_D$ total number, 'paradiamagnetism', and so forth, which differentiate the latter from the closed-shell ones. The scene of the SOC manifesration is provided with either odd electrons, which do not participate in the formation of covalent bonds of the molecule, or electrons that are withdrawn from the covalent bonds in due course of their stretching (towards dissociation). In both cases spacing between the electrons should exceed a critical value $R_{crit}$ characteristic for the covalent bond under consideration [2]. The two cases can be conventionally attributed to static and dynamic open-shell molecules. Molecules with multiple covalent bonds, such as well known $sp^2$-nanocarbons (fullerenes, carbon nanotubes, and graphene), are an example of the first type of the molecules [3]. All the outlined molecules are characterised by equilibrium C-C distances, a predominant majority of which exceeds $R_{crit}$ of 1.395Å characteristic for closed-open shell transition provided with double C-C bonds [2]. As for the second type molecules, they cover a very large area of species under stretching that concerns this or that covalent bond. The current paper is devoted to the consideration of both static and dynamic $sp^2$-carbonaceous open-shell molecules aimed at determination of such SOC parameters as SOC energy $E_{SO}$ and constant $a_{SO}$ as well as *g*-factors Lande. As shown in [1], unrestricted Hartree-Fock (UHF) approach quite efficiently discloses the SOC in light-element molecules. Data, presented below, were obtained by using semi-empirical AM1 UHF version implemented in the CLUSTER-Z1 codes [4].

## 2. General concepts

Spin–orbit interaction Hamiltonians are most effectively derived by reducing the relativistic four-component Dirac–Coulomb–Breit operator to two components and separating spin-independent and spin-dependent terms. This reduction can be achieved in many different ways one of which leads to the most renowned Breit-Pauli spin–orbit Hamiltonian related to many-electron systems given below in the form in which the connection to the Coulomb potential and the symmetry in the particle indices is apparent [5].

$$\widehat{H}_{SO}^{BP} = \frac{e^2\hbar}{2m^2c^2}\sum_i\left(-\boldsymbol{\nabla}_i\left(\sum_I\left\{\frac{Z_I}{r_{iI}} - \sum_{j\neq i}\frac{1}{r_{ij}}\right\}\right) \times \mathbf{p}_i\right)\cdot\hat{\mathbf{s}}_i \ . \quad (1)$$

Here $I$ labels nuclei while $i$ and $j$ do the same for electrons, $Z_I$ is the charge of nucleus $I$, $\hat{\boldsymbol{p}}_i$ and $\hat{\mathbf{s}}_i$ are pulse and spin operators of the $i$th electron, respectively, $e$ and $m$ are the charge and the mass of an electron and $c$ is the speed of light. Terms in curly brackets present a self-consistent potential for the $i$th electron of atomic centre $I$ $U_I(r_i)$ where the first summand describes the interaction of the spin magnetic moment of the electron with the magnetic moment that arises from its orbiting in the field of the nucleus $I$. The second covers both spin- same- and spin-other-orbit parts related to two-electron Coulomb interaction.

The problem how to compute spin-orbit effects has a long story. As shown (see review [6, 7] for details), four-component *ab initio* methods automatically include scalar and magnetic relativistic corrections, but they put high demands on computer resources. The two-component methods, based on one of the forms of $\widehat{H}_{SO}^{BP}$ Hamiltonians, treat SOC either perturbationally or variationally. Most of these procedures start off with orbitals optimized for a spin-free Hamiltonian (typical to closed-shell molecules). Spin–orbit coupling is added then at a later stage [5]. Just recently a new approach has been suggested [8] based on the idea that the SOC consideration mandatory requires one of the forms of a multireference CI theory (MRCI) due to which the computation of spin-orbit effects can be introduced in the body of one of the MRCI tools. It implies that the reference functions are obtained in the framework of UHF calculations while matrix elements of the $\widehat{H}_{SO}^{BP}$ Hamiltonian are expressed in the Fock-matrix-like form to be adapted further for calculation of spin-orbit integrals. Attention should be drawn to the decisive role of the UHF approach and open-shell character of the studied molecular object. Unfortunately, the study, aimed at the calculation of splitting the molecules ground state levels [8], does not contain a comparative analysis of the data followed from the UHF solution only and those, based on the SOC integrals calculations, which does not allow to judge how much the latter was essential. On the other hand, a great deal of arguments, presented in [1], gives a convincing confirmation of the UHF ability to get a correct snapshot of the SOC effects. In what follows we shall use the UHF data from the standard output file to evaluate such important SOC characteristics as the SOC constant $a_{SO}$ and Lande factor $g$.

The atom potential $U_I(r_i)$ in eq. (1) is spherically symmetric, due to which $\boldsymbol{\nabla}_i U_I(r_i) = \left(\frac{d}{dr_i}U_I(r_i)\right)\cdot\frac{\mathbf{r}_i}{r_i}$ . Assuming that the interatomic interaction does not affect the potential symmetry much, the spin-orbit Hamiltonian for a many-electron system can be written in the form:

$$\begin{aligned}\widehat{H}_{SO}^{BP} &= \frac{\hbar}{2m^2c^2}\sum_I\sum_i\left(\boldsymbol{\nabla}_i U_I(r_i)\right)\times\hat{\mathbf{p}}_i\cdot\hat{\mathbf{s}}_i \\ &= \frac{\hbar}{2m^2c^2}\sum_I\sum_i\left(\left(\frac{d}{dr_i}U_I(r_i)\right)\cdot\frac{\mathbf{r}_i}{r_i}\right)\times\hat{\mathbf{p}}_i\cdot\hat{\mathbf{s}}_i \quad (2)\\ &= \frac{\hbar}{2m^2c^2}\sum_I\sum_i\left(\left(\frac{d}{dr_i}U_I(r_i)\right)\cdot\frac{1}{r_i}\right)\hat{\boldsymbol{l}}_i\cdot\hat{\mathbf{s}}_i.\end{aligned}$$

Averaging the potential over eigenfunctions related to the total orbital angular momentum $\hat{\mathbf{L}}_I = \sum_i\hat{\boldsymbol{l}}_i$ gives us

$$\hat{H}_{SO}^{av} \approx \frac{\hbar^2}{2m^2c^2} \sum_I \overline{\left\{\frac{1}{r_\iota}\left(\frac{d}{dr_\iota}U_I(r_\iota)\right)\right\}} \cdot \hat{L}_I \cdot \sum_i \hat{s}_i$$

$$= \frac{\hbar^2}{2m^2c^2} \sum_I \overline{\left(\frac{1}{r_\iota}\left(\frac{d}{dr_\iota}U_I(r_\iota)\right)\right)}\bigg|_{L_I} \cdot \hat{L}_I \cdot \hat{S}_I. \tag{3}$$

The eigenvalues of the Hamiltonian $H_{SO}^{av}$, once rewritten as

$$\hat{H}_{SO}^{av} = \sum_I a_{SOI}(\hat{L}_I \cdot \hat{S}_I), \tag{4}$$

have the form

$$\mathcal{E}_{SO} = \frac{1}{2}\sum_I a_{SOI}(J_I(J_I+1) - L_I(L_I+1) - S_I(S_I+1)), \tag{5}$$

Here

$$a_{SOI} = \frac{\hbar^2}{2m^2c^2} \overline{\left(\frac{1}{r_\iota}\left(\frac{d}{dr_\iota}U_I(r_\iota)\right)\right)}\bigg|_L. \tag{6}$$

For regular states $a_{SO} > 0$ due to which the ground term corresponds to the minimum value of the total angular momentum $J = |L - S|$.

Addressing to open-shell molecules, in terms of the UHF output data Eq. 6 can be approximated as

$$a_{SO} \approx \frac{\hbar^2}{2m^2c^2}\left(\frac{1}{R}\left(\frac{d}{dR}E_{SO}(R)\right)\right), \tag{7}$$

where $E_{SO}(R)$ is the part of the molecule total energy dependent on the length of the covalent bond that promotes the closed-open shell transformation and is responsible for SOC. How correct evaluation of the $a_{SO}$ can be achieved under such approximation will be demonstrated for the stretched ethylene molecule in the next Section.

## 3. Dynamic open-shell molecules: SOC in stretched ethylene

Main UHF characteristics, which accompany stretching the ethylene C-C bond up to 2Å, are shown in Fig. 1a. Equilibrium C-C distances constitute 1.326Å and 1.415Å in singlet and triplet states of the molecule, respectively. As seen in the figure, as stretching of the bond increases, energies $E_{sg}(R)$ and $E_{tr}(R)$ approach each other up to quasidegeneracy, which is characteristic for biradicals and which is necessary for an effective SOC [5]. Simultaneously, the number of effectively unpaired electrons $N_D$, which is zero until C-C distance reaches $R_{crit}$, starts to grow manifesting a gradual radicalization of the molecule as the bond is stretched as well as exhibiting the transformation of the molecule behavior from closed-shell to open-shell one when $R_{crit}$ is overstepped.

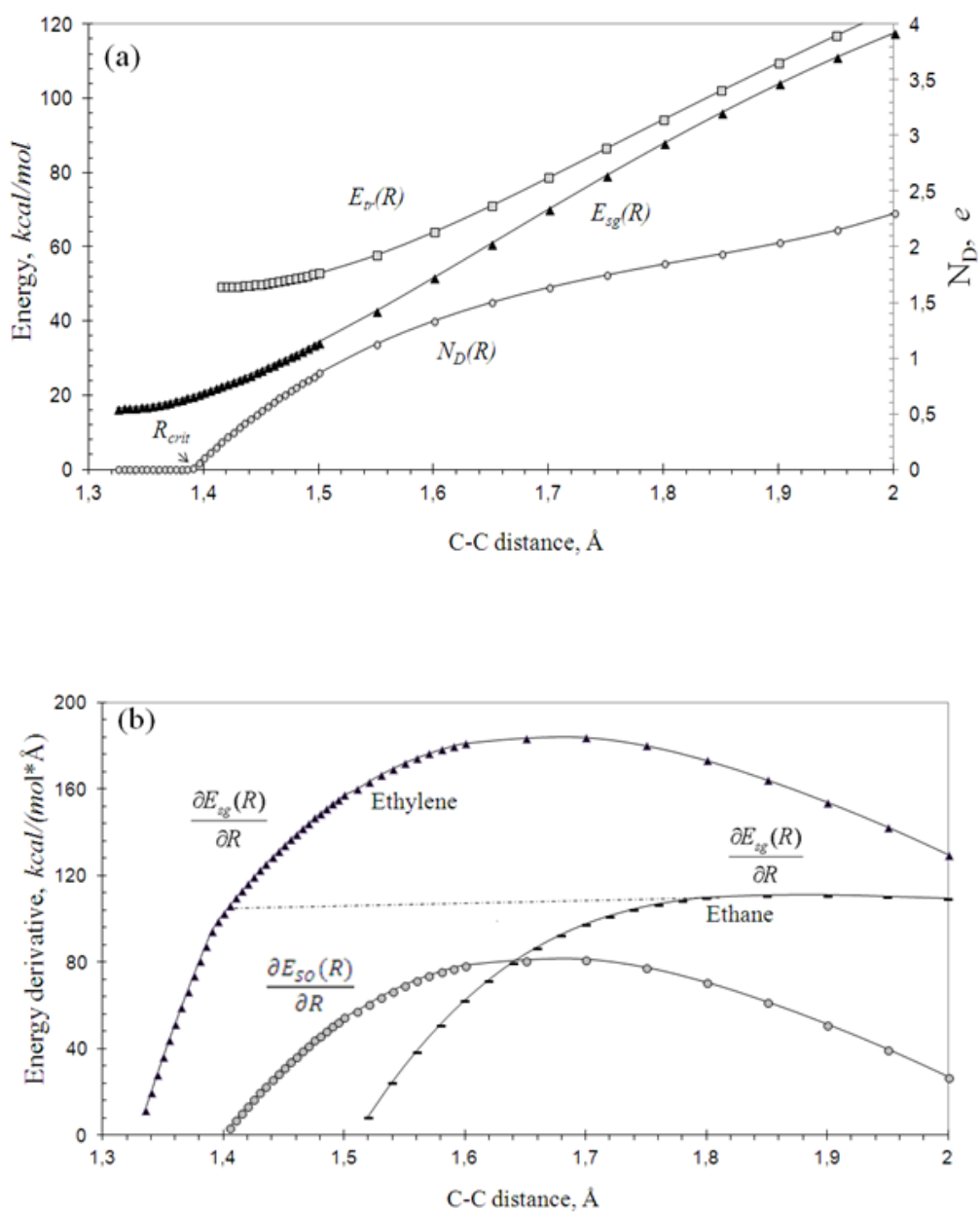

**Figure 1.** a. Energy of singlet and triplet states as well as the total number of effectively unpaired electrons $N_D$ of ethylene vs C-C distance. b. Energy derivatives vs C-C distance (see text) (UHF calculations).

The force applied to the ethylene C-C bond under stretching is presented in Fig. 1b. At the beginning it proceeds linearly starting, however, to slow down when $R$ is approaching $R_{crit}$. A clearly seen kink is vivid in the region. For comparison, the curve with horizontal bars presents the force caused by stretching a single C-C bond of ethane, $R_{crit}$ for which constitutes 2.11Å [2], that is why the molecule remains closed-shell one within the interval of C-C distances presented in the figure. As seen, the force is saturated at the level of 110 kcal/(mol*Å) that is close to the kink position on the ethylene curve. It is quite reasonable to suggest that the force excession over this value in the latter case is caused by the closed-open shell transformation of the ethylene molecule over 1.4Å due to which the force excess is caused by the SOC.

Consequently, this excess can be considered as $dE_{SO}(R)/dR$ that is presented by the gray-ball curve in the figure. Using the curve values in the C-C distance interval from 1.40Å to 1.47Å, which is typical for the C-C bond dispersion in fullerenes, carbon nanotubes, and graphene, and substituting them into Eq. 7, one can obtain the $a_{SO}$ constant laying in the interval from 15 meV to 110 meV, which is typically expected for molecules of light elements [5] and which was predicted [9] and experimentally determined [10] for graphene. It is necessary to point out that the $dE_{SO}(R)/dR$ force maximum amplitude for ethylene molecule is well consistent with those determined under uniaxial deformation of benzene molecule and graphene [11] due to which the outlined $a_{SO}$ constant values can be considered as typical for the whole family of $sp^2$ nanocarbons.

## 4. Static open-shell molecules: g-factor Lande

One of the most convincing evidence of the connection of the open-shell nature of fullerene $C_{60}$ with SOC are amazing features of its behavior in a magnetic field [12, 13]. Providing the kind permission of S.V.Demishev, Fig. 2 reproduces the latter related to the molecules in condensed state. Given below is a new view on the data suggested in the current paper from the viewpoint of SOC in open-shell molecules.

As seen in panel a, the molecules, once exhibiting diamagnetic behavior up to T=40 K, change the latter for paramagnetic one below the temperature. The observed 'paradiamagnetism' (in terms of Margarill and Chaplik [14]) is typical for the SOC phenomenology directly evidencing the spin mixing. Shown in panel b demonstrates that paramagnetic magnetization has a standard dependence on the magnetic field but with the Lande g-factor lying between 1 and 0.3. The strong reduction of the factor clearlyr evidences that electrons responsible for the observed magnetization are bound. The most amazing finding is shown in panel c, demonstrating the availability of three g-factor values revealed in the course of magneto-optical study at pulsed magnetic field up to 32 T in the frequency range $\nu$ = 60−90 GHz at $T$ = 1:8K. If two first features, namely, paradiamagnetism and deviation of the g-factor values from those related to free electrons, were observed in other cases as well, the third one is quite unique and is intimately connected with the electron structure of fullerene $C_{60}$.

Related to paramagnetic part, the feature should be attributed to a peculiar behavior of the molecule odd $p_z$ electrons while the other three valence electrons of each carbon atom are $sp^2$-configured and involved in the formation of spin-saturated σ bonds. Due to open-shell character of the molecule, part of the odd electrons, of the total number 9.87$e$, are unpaired and fragmentarily distributed over the molecule atoms. A color image of the molecule shown in Fig. 3a gives a view of this distribution. The relevant spin density on each of the 60 atoms is given by histogram. The total spin density equals zero while its negative and positive values are symmetrically distributed. As seen in the figure, the distribution clearly reveal well configured compositions of the odd electrons ('local spins' [15]) with identical positive and negative spin-density values within the latter. Going over the distribution from the highest to the lowest density, one can distinguish two hexagons, 12 singles and three sets of pairs covering 12 atoms each. Evidently, these configurations will differently respond to the application of magnetic field thus mimicking the behavior of a molecule with different atoms. This allows for exploring the atomic expression for the Lande g-factor

$$g = 1 + \frac{J(J+1) - L(L+1) + S(S+1)}{2J(J+1)} \qquad (8)$$

to evaluate g-values related to each set of the local spin configurations of the molecule.

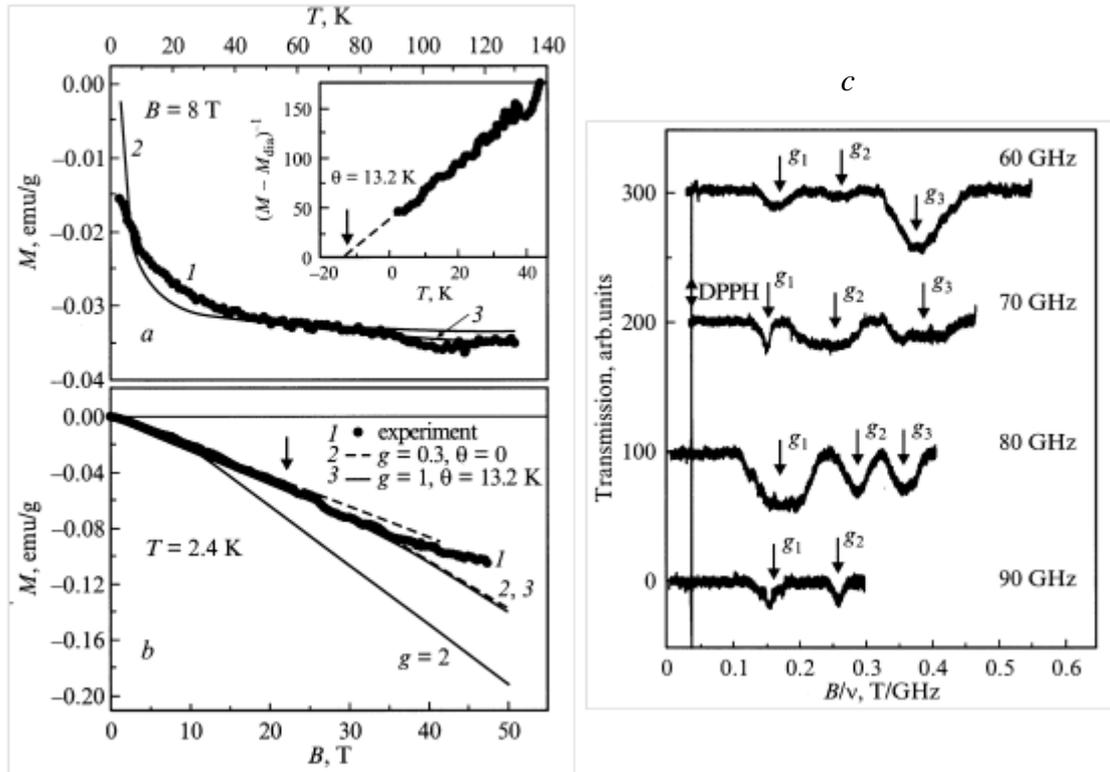

**Figure 2**. Temperature (*a*) and field (*b*) dependence of magnetization for $C_{60}$-TMTSF-$2CS_2$ molecular complex. *1* . experimental data for *M(T)* and *M(B)*, *2* and *3* . simulations of *M(T)* and *M(B)* (see details in [14]). Inset in part (*a*) shows temperature dependence of magnetization in coordinates $(M - M_{dia})-1 = f(T)$. c. ESR absorption lines in $(ET)2C_{60}$ molecular complex at $T = 1.8K$ [14] (by kind permission of S.V.Demishev).

Application of magnetic field disturbs the antiferromagnetic regularity of the outlined localspins towards ferromagnetic one (see Fig. 3b) that corresponds to the maximum of magnetic ordering. If experimentally observed, the magnetic properties of fullerene molecules should reflect a variability of g-factor caused by different configuration sets of spins. The corresponding values of the factor can be evaluated by the following way.

1. *Hexagon of upright local spins*.

The total spin $S = 3$, the total orbital momentum of the hexagon $p_z$ electrons $L = 6$. Consequently, the total angular momentum $J$ of the hexagon has seven components: (6+3), 8, 7, 6, 5, 4, (6-3). According to Eq. 5, the lowest energy state corresponds to the minimum value of the total angular momentum $J = 3$. Following Eq. 9, the relevant Lande factor $g_6=0.25$.

2. *Pairs of upright local spin*.

The total spin $S = 1$, the total orbital momentum of pair $p_z$ electrons is $L = 2$ and the total angular momentum $J$ has three components: (1+2), 2, (2-1). Taking the least value component, obtain the relevant Lande $g_2=0.5$

3. *Local spin singles*. For a single $p_z$ electron, the total spin $= \frac{1}{2}$, the orbital angular momentum $L = 1$, while the total angular momentum $J$ has two components: 3/2 and 1/2. According to eq. (6), the lowest energy term corresponds to $J = ½$ so that the Lande factor $g_1=0.66$.

Thus obtained g values are collected in Table 1 alongside with experimental data. As seen in the table, experimental $g_1$, $g_2$, and $g_3$ can be attributed to singles, pairs, and hexagons of local spins, respectively. The calculated and experimental values are in a good consent. The absence of exact coincidence is evident due to a few reasons among which two next are the most important. 1) Eq. 9 is too simplified; its application allowed exhibiting quasi-polyatomic structure of fullerene molecule related to different configurations of local spins while the exact g values determination is much more complicated problem. 2) The evaluation of g-factors was performed for ferromagnetic ordering of local spins in $C_{60}$. Experimentally, the studied crystals showed 'paradiamagnetism' exhibiting above T=40K diamagnetic behaviour that is continued as paramagnetic one below the temperature. Dominating diamagnetism is definitely provided with σ electrons that are in majority. Local spins related to unpaired $p_z$ electrons can be seen on the diamagnetic temperature-stable background only due to specific temperature dependence. Realization of the 1/T paramagnetic dependence in practice is caused by free rotation of ferromagnetically spin-configured molecules occurred at very low temperatures [15]. Actually, when such a rotation is forbidden, as it is in the case of narrow graphene ribbons (molecules) terminated by hydrogen atoms and rigidly fixed with respect to immobile substrate [16], the ferromagnetic behaviour of the molecules, having the same explanation as was suggested above for fullerene $C_{60}$, has been clearly observed.

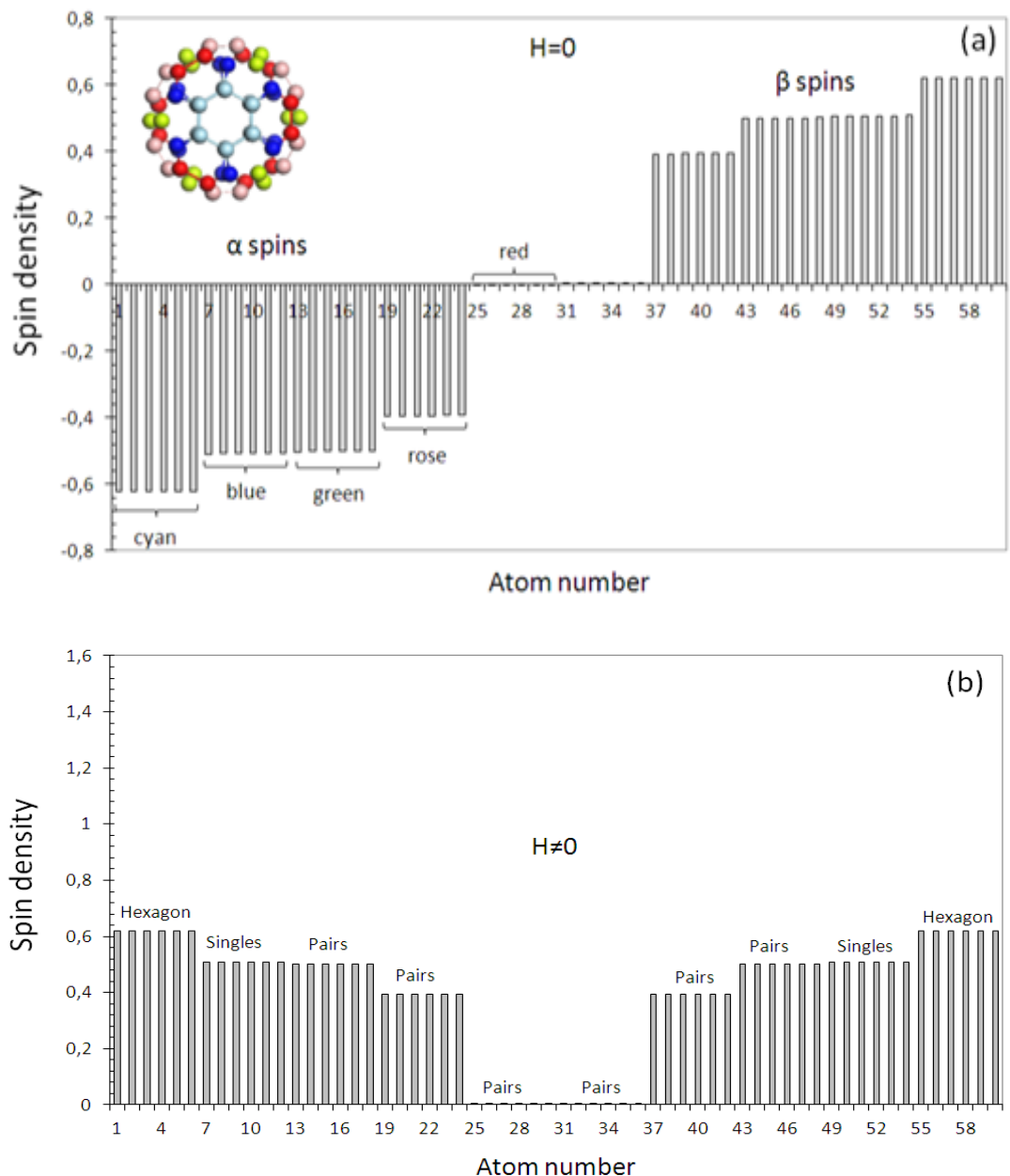

**Figure 3**. Spin density distribution over $C_{60}$ atoms in the absence (a) and presence (b) of magnetic field. Insert exhibits the color image of the local spin distribution over the molecule atoms following notations in panels *a* and *b* (UHF calculations).

**Table 1**. g-Factors of fullerene $C_{60}$

| Calculated | | Experimental [13] | |
| --- | --- | --- | --- |
| Attribution | Value | Attribution | Value |
| Hexagons | 0.25 | $g_3$ | 0.19 ± 0.01 |
| Singles | 0.50 | $g_2$ | 0.27± 0.02 |
| Pairs | 0.66 | $g_1$ | 0.43 ±0.03 |

### 4. Conclusion

The performed study has shown that a spin-orbit concept of open-shell molecules [1] finds confirmation in the case of $sp^2$ nanocarbons. Dynamic and static differentiation of the molecules occurred useful for their properties to become clear and computationally feasible. The evaluated SOC constants $a_{SO}$ are well consistent with empirically determined characteristics of the Rashba-Sheka effect [17] in graphene [10]. In view of SOC, the 'paradiamagnetic' behaviour of crystalline fullerene $C_{60}$ as well as drastic reduction and splitting of its g factor [12, 13] receive a comprehensive explanation in terms of local spins and varying configurations of the latter. The UHF computational scheme has shown itself able to adequately reproduce spin-orbit effects. Just recently has been shown [18] that peculiarities of the UHF formalism can be exactly reproduced within the framework of relativistic molecular theory thus emphasizing their spin-orbit origin.

**Acknowledgements**

The authors are grateful to I. Mayer, J. Karwowski, S.Vinitskii, A. Gusev, Yu. P. Rybakov for fruitful discussions and S.V.Demishev for kind permition to use experimental data at our discretion. The work was performed under financial support of the Peoples' Friendship University of Russia, grant: 022203-0-000.